\documentclass[aps,pra,twocolumn,showpacs,superscriptaddress,floatfix,nofootinbib]{revtex4}
\usepackage{graphicx}
\usepackage{nicefrac}
\usepackage{amsmath}
\usepackage{amsfonts}
\usepackage{amssymb}
\usepackage{amsthm}
\usepackage{epsf}
\usepackage{bm}
\usepackage{bbm}
\usepackage{longtable}

\usepackage{dcolumn}

\newcolumntype{.}{D{x}{}{-1}}
\newcommand{\bsigma}{\bm{\sigma}}
\newcommand{\balpha}{\bm{\alpha}}

\newcommand{\bfr}{\bm{r}}
\newcommand{\bfk}{\bm{k}}
\newcommand{\bfp}{\bm{p}}

\newcommand{\bfx}{\bm{x}}

\newcommand{\bfu}{\bm{u}}

\newcommand{\hr}{\hat{\bfr}}

\newcommand{\hp}{\hat{\bfp}}

\newcommand{\Za}{{Z\alpha}}

\newcommand{\vare}{\varepsilon}

\newcommand{\SixJ}[6]{
        \left\{
        \begin{array}{ccc}
        #1  & #2  & #3 \\
        #4  & #5  & #6 \\
        \end{array}
        \right\}
        }
\newcommand{\NineJ}[9]{
        \left\{
          \begin{array}{ccc}
            #1  & #2  & #3 \\
            #4  & #5  & #6 \\
            #7  & #8  & #9 \\
          \end{array}
        \right\} }

\newcommand{\Dmatrix}[4]{ \left(
          \begin{array}{cc}
            #1  & #2   \\
            #3  & #4   \\
          \end{array}
        \right) }

\def\ketm#1{  \left\vert  #1   \right\rangle   }

\begin{document}

\title{Electron-atom bremsstrahlung: double differential cross section and
polarization correlations}

\author{Vladimir A. Yerokhin}
\affiliation{Institute of Physics, University of Heidelberg, Philosophenweg
  12, D-69120 Heidelberg, Germany}
\affiliation{Gesellschaft f\"ur Schwerionenforschung, Planckstra{\ss}e 1,
D-64291 Darmstadt, Germany}
\affiliation{Center for Advanced Studies, St.~Petersburg State
Polytechnical University, Polytekhnicheskaya 29,
St.~Petersburg 195251, Russia}

\author{Andrey Surzhykov}
\affiliation{Institute of Physics, University of Heidelberg, Philosophenweg
  12, D-69120 Heidelberg, Germany}
\affiliation{Gesellschaft f\"ur Schwerionenforschung, Planckstra{\ss}e 1,
D-64291 Darmstadt, Germany}

\begin{abstract}

The leading-order electron-atom
bremsstrahlung is investigated within the rigorous relativistic approach based on the
partial-wave representation of the Dirac wave functions in the external
atomic field. Approximating the atomic target by an effective local
potential, we calculate the double-differential cross section and
the polarization correlations in a wide range of the
impact energies. Connection between the bremsstrahlung at the
hard-photon end point of the spectrum and the continuum-threshold limit of the
radiative recombination is studied. A detailed analysis of the screening
effect and the energy dependence of the polarization correlations is presented,
with the main focus on the high impact energy region.

\end{abstract}


\maketitle

\section{Introduction}

Bremsstrahlung, the emission of a photon by an electron scattering from an
atom, is one of the fundamental processes that occur in the electron-atom
collisions.
For a large region of the impact energies of the incoming electron, the atom
can be well represented by a static central (screening) potential, thus
ignoring the exchange interaction between the
incoming electron and the target and virtual excitations
of the target. The corresponding approximation is sometimes referred as the ``ordinary''
bremsstrahlung mechanism, to distinguish from the so-called ``polarization''
bremsstrahlung, in which the excess energy is transferred to the
target and the photon is emitted from the atomic core. The polarization
bremsstrahlung is the dominant mechanism in, e.g.,  the {\it proton}-atom
collisions (see, e.g., Ref.~\cite{korol:01}), where
the ordinary bremsstrahlung is suppressed by the small electron-to-proton mass
ratio. In the electron-atom collisions, however, the polarization mechanism is
usually not significant for the not-too-small energies of the incoming
electron.

Within the screening-potential approximation, the bremsstrahlung process can
be described rigorously (to the leading order in the fine-structure constant
$\alpha$) within the exact relativistic approach based on the
partial-wave representation of the Dirac continuum electron states with a
fixed value of the asymptotic momentum. This approach is clearly preferable as
compared to numerous approximate treatments reported in the literature but it
is also more difficult for practical implementations. The
partial-wave decomposition of the initial and final electron states, together
with the multipole expansion of the wave function of the emitted photon,
lead to a large number of expansion terms that have to be summed up
until convergence is reached. Despite technical difficulties
encountered, the first partial-wave expansion calculations
were reported already in 1960s
\cite{rozics:64,brysk:69}. The first accurate numerical results
were obtained by Tseng and Pratt in 1970s. In a series of
calculations \cite{tseng:71,tseng:74:prl,lee:77,tseng:79:a}, they reported
results for the single- and
double-differential cross sections for a wide interval of the
impact energies. Polarization
correlations in the double-differential cross section were studied in
Ref.~\cite{tseng:73}. More recently, several calculations of the triple-differential cross
section and the corresponding polarization correlations
were performed \cite{schaffer:96,keller:97,tseng:02}.

Experimental investigations of the electron-atom bremsstrahlung, numerous in
1960s and 1970s \cite{nakel:94}, became rather
sparse during the following decades. In the
last years, however, advent of a new generation of the Compton polarimeters
\cite{tashenov:06} revived the experimental interest to
bremsstrahlung. In particular, it became possible to perform experiments
with the spin-polarized electrons, which opens up new vistas for
experimental studies of various correlations between the polarizations of
the incoming electron and the emitted photon. Such experiments are currently
underway in Gesellschaft f\"ur Schwerionenforschung (GSI) \cite{tashenov:unpublished}
and Technical University of Darmstadt \cite{maertin:CAARI}.

Because of the experimental limitations in the past, the
previous bremsstrahlung calculations were focused mainly on the cross
sections. The only detailed partial-wave expansion study of the
polarization correlations
was accomplished by Tseng and Pratt \cite{tseng:73}. Detailed
as it is, this calculation is not sufficient to cover all present experimental
needs. In particular, the region of the impact energies
$E>500$~keV, presently available for the experimental investigation in GSI,
has not been carefully investigated in that work.
Besides this, the results reported in Ref.~\cite{tseng:73} have never been carefully
checked by an independent calculation. In the present investigation, we aim to
cover this gap by checking and extending the previous bremsstrahlung results by Tseng
and Pratt.

The paper is organized as follows. In Sec.~\ref{sec:theory}, we present
relativistic formulas for the double-differential
bremsstrahlung cross section and polarization
correlations, obtained within the density-matrix
formalism. Sec.~\ref{sec:numerics} describes the numerical approach used in
this work. In Sec.~\ref{sec:tip} we discuss the connection between the
bremsstrahlung at the hard-photon end point of the spectrum and the
continuum-threshold limit of the radiative recombination. Numerical results
are presented and discussed in Sec.~\ref{sec:num}.

The relativistic units ($\hbar=c=1$) are used throughout this paper.

\section{Theory}
\label{sec:theory}

In the present investigation we consider the bremsstrahlung of an electron
scattering from an atom, which is represented by a static central potential.
The scattered electron in the final state is assumed to be not observed.
The kinematics of the process is defined as follows. The reference frame is
the rest frame of the atom. The $z$ axis is directed along the asymptotic
momentum of the incident electron $\bfp_i$. The $xz$ plane (also referred to
as the {\em reaction} plane) is defined by $\bfp_i$ and
the momentum of emitted photon $\bfk$. In this frame, the direction of the
emitted photon is defined just by the polar angle $\theta_k$, $\cos \theta_k =
{\hat{\bfp}}_i\cdot \hat{\bfk}$, where $\hat{\bfx}\equiv \bfx/|\bfx|$.

In order to study the polarization correlations between the incident
electron and the emitted photon, we express the density matrix of the final
state $\bigl< \bfk \lambda| {\rho}_f |\bfk \lambda^{\prime}\bigr>$\footnote{
More exactly, this
is the reduced density matrix of the emitted photons, with the quantum numbers
of the (unobserved) final-state electron traced out. In the context of the
present paper, we refer to it just as the density matrix of the final state.
}
in terms of the initial-state
density matrix as
\begin{align} \label{1}
\bigl< \bfk \lambda| {\rho}_f |\bfk \lambda^{\prime}\bigr>
 &\  =
  \sum_{m_i m^{\prime}_i m_f} \int d\Omega_f\,
    \bigl< \bfp_i m_i| \balpha\cdot \hat{\bfu}_{\lambda}\, e^{i\bfk\cdot\bfr}
         |\bfp_f m_f\bigr>^*\,
 \nonumber \\ &\times
    \bigl< \bfp_i m^{\prime}_i| \balpha\cdot \hat{\bf u}_{\lambda^{\prime}}\, e^{i\bfk\cdot\bfr}
         |\bfp_f m_f\bigr>\,
 \bigl< \bfp_i m_i|{\rho}_i|\bfp_i m_{i}^{\prime}\bigr>\,,
\end{align}
where $\bfp_i$, $m_i$ ($m_i^{\prime}$) and $\bfp_f$, $m_f$
are the asymptotic momentum and the spin projection
of the incident and scattered electron states, respectively;
$\bfk$ and $\lambda$ ($\lambda^{\prime}$) are the momentum and the helicity
of the emitted photon ($\lambda = \pm 1$), $\hat{\bfu}_{\lambda}$ is the unit
polarization vector of the photon, and $\Omega_f$ is the solid angle of the
scattered electron. The energy of the emitted photon $k \equiv |\bfk|$ is fixed by
$
 k  = \vare_i-\vare_f\,,
$
where $\vare_n = \sqrt{p_n^2+m^2}$.

The advantage of the present formulation is that
the density matrix (\ref{1}) contains the full information about the polarization
properties of the emitted photon (see, e.g., Ref.~\cite{balashov:book}),
\begin{align} \label{2}
\bigl< \bfk \lambda| {\rho}_f |\bfk \lambda^{\prime}\bigr>
 = \frac12 \, {\rm Tr}\left[ {\rho}_f \right]\,\Dmatrix{1+P_3}{P_1-iP_2}{P_1+iP_2}{1-P_3}  \,,
\end{align}
where $P_i$ are the Stokes parameters. We note that the sign of the Stokes
parameters depends on the definition of the
circular polarization unit vectors, see Eq.~(\ref{11}) below and the text after it.
The trace of
the density matrix is, up to a prefactor, the double-differential cross
section summed over all photon polarizations.
Following Ref.~\cite{tseng:71}, we introduce the normalized cross section
$\sigma(k)$ as
\begin{align} \label{3}
 \sigma(k) \equiv &\
\frac{k}{Z^2}\, \frac{d\sigma}{dk} =
2\pi\, \int_{-1}^{1} d(\cos \theta_k)\,\frac{k}{Z^2}\,
\frac{d\sigma}{dk\,d\Omega_k}
 \nonumber \\  &
= \frac{1}{32\pi}\, \frac{k^2}{p_i^2}\, \frac{\alpha}{Z^2}\,
\int_{-1}^{1} d(\cos \theta_k)\,
  {\rm Tr}\left[ {\rho}_f \right]\,,
\end{align}
where $Z$ is the nuclear charge of the target atom and the continuum
electron wave function is assumed to be normalized on the energy scale.

The Stokes parameters $P_1$ and $P_2$
describe the linear polarization of the emitted photon. They can be
determined experimentally by measuring the intensities $I_{\phi}$ of the
linearly polarized photon emission at different angles $\phi$ with respect to
the reaction plane,
\begin{align} \label{4}
 P_1 &\ =
 \frac{I_{0^{\circ}}-I_{90^{\circ}}}{I_{0^{\circ}}+I_{90^{\circ}}}\,,
 \\\
 P_2 &\ =
 \frac{I_{45^{\circ}}-I_{135^{\circ}}}{I_{45^{\circ}}+I_{135^{\circ}}}\,.
 \label{4a}
\end{align}
Instead of the $P_1$ and $P_2$ parameters, it is often convenient to describe
the linear polarization by the polarization ellipse in the plane perpendicular
to the photon momentum $\bfk$. The ellipse is defined by the relative length
of the principal axis $P_L$ which reflects the degree of the linear
polarization,
\begin{align} \label{5}
P_L = \sqrt{P_1^2+P_2^2} \,,
\end{align}
 and the tilt angle $\chi$,
\begin{align} \label{6}
\chi = \frac12\,{\rm acrtan}\frac{P_2}{P_1}\,.
\end{align}
The Stokes parameter $P_3$ defines the degree of the circular polarization of the
emitted photon,
\begin{align} \label{7}
P_3 = \frac{W(+1)-W(-1)}{W(+1)+W(-1)}\,,
\end{align}
where $W(+1)$ and $W(-1)$ are weights of the right- and left-polarized
photons, respectively.

We now turn to the evaluation of Eq.~(\ref{1}).
The initial-state density matrix is represented in terms of
the spherical tensors $\rho^{(i)}_{\kappa}$ of rank $\kappa = 0$ and 1
as \cite{balashov:book}
\begin{align} \label{8}
 \bigl< \bfp_i m_i|{\rho}_i|\bfp_i m_{i}^{\prime}\bigr>
 = \sum_{\kappa q} (-1)^{\nicefrac{1}{2}-m_i^{\prime}}\, C_{\nicefrac{1}{2} m_i\,,
   \nicefrac{1}{2}\, -m_i^{\prime}}^{\kappa q} \, \rho^{(i)}_{\kappa q}\,.
\end{align}
The components of $\rho^{(i)}_{\kappa}$ are expressed in terms of the
polarization vector of the incoming electron $P_i = (P_x,P_y,P_z)$ as
\begin{align} \label{9}
\rho^{(i)}_{00} = \frac1{\sqrt{2}}\,, \ \
\rho^{(i)}_{10} = \frac1{\sqrt{2}}\,P_z\,, \ \
\rho^{(i)}_{1\,\pm 1} = \mp \frac12\,\bigl(P_x\mp i P_y)\,.
\end{align}
The spherical-wave decomposition of the
ingoing ($+$) and outgoing ($-$) continuum electron wave function with a
definite asymptotic momentum is \cite{eichler:07:review}
\begin{align} \label{10}
|\bm{p} m\bigr>  = 4\pi\, \sum_{\kappa\mu} i^l\, e^{\pm i\Delta_{\kappa}}\,C^{j\mu}_{l m_l, \frac12
  m}\, Y_{lm_l}^*(\hp)\,   |\vare \kappa \mu\bigr>
\,,
\end{align}
where $j = |\kappa|-1/2$, $l = |\kappa+1/2|-1/2$,
$\Delta_{\kappa} = \sigma_{\kappa}+(l+1)\,\pi/2$, and $\sigma_{\kappa}$ is the
asymptotic phase of the wave function, see
Eq.~(\ref{10b}) below. The function $|\vare \kappa \mu\bigr>$ is the Dirac
eigenstate with the energy $\vare = \sqrt{p^2+m^2}$,
the relativistic angular quantum number $\kappa$, and the
angular momentum projection $\mu$, represented by
\begin{align} \label{10a}
|\vare \kappa \mu\bigr> = \left(
        \begin{array}{c}
          g_{\vare,\kappa}(r)\, \chi_{\kappa\mu}(\hr) \\
          i f_{\vare,\kappa}(r)\, \chi_{-\kappa\mu}(\hr)
        \end{array}
\right) \,,
\end{align}
where $g$ and $f$ are the upper and lower radial components, respectively, and
$\chi_{\kappa\mu}$ are the spherical spinors \cite{rose:61}. The wave function
is normalized on the energy scale and its asymptotic (as $r\to\infty$) behavior
is
\begin{align} \label{10b}
g_{\vare,\kappa}(r) \simeq \frac1r\, \left( \frac{\vare+m}{\pi
  p}\right)^{1/2}\, \cos \bigl[pr+\sigma_{\kappa}+\eta \ln(2pr)\bigr]\,,
\end{align}
where $\eta = Z_{\infty}\alpha \vare/p$, and
$Z_{\infty}$ is the effective nuclear charge of the atom at large distances,
$Z_{\infty} = \lim_{r\to\infty} r V_{\rm scr}(r)$, with $V_{\rm scr}$ being
the atomic potential.

The spherical-wave decomposition of the photon field with the helicity
(circular polarization) $\lambda= \pm 1$ is
\begin{align} \label{11}
\hat{\bfu}_{\lambda}\, e^{i\bfk\cdot\bfr} =
 \sqrt{2\pi}\, \sum_{LMp} i^L\,\sqrt{2L+1}\, (i\lambda)^p \,
  \bm{a}^{(p)}_{LM}(\hat{\bfr})\, D^L_{M\lambda}(\hat{\bm{k}})\,,
\end{align}
where the components of the
circular-polarization vector $\hat{\bfu}_{\lambda}$
are defined as \cite{rose:61}  $u_{1} = (u_x + i u_y)/\sqrt{2}$
and $u_{-1} = (u_x - i u_y)/\sqrt{2}$,
$D^L_{M\lambda}$ is Wigner's $D$ function \cite{varshalovich},
$\bm{a}^{(p)}_{LM}$ are the magnetic ($p =0$) and electric ($p=1$) vectors defined by
\begin{align} \label{12}
 \bm{a}^{(0)}_{LM}(\hat{\bfr}) &\ = j_L(kr)\,\bm{Y}_{LLM}(\hat{\bfr})\,, \\
 \bm{a}^{(1)}_{LM}(\hat{\bfr}) &\ = j_{L-1}(kr)\,\sqrt{\frac{L+1}{2L+1}}\,\bm{Y}_{L\,L-1\,M}(\hat{\bfr})
 \nonumber \\ &
- j_{L+1}(kr)\,\sqrt{\frac{L}{2L+1}}\,\bm{Y}_{L\,L+1\,M}(\hat{\bfr})\,,
\end{align}
$\bm{Y}_{JLM}$ is the vector spherical harmonics \cite{varshalovich},
\begin{align} \label{13}
\bm{Y}_{JLM}(\hat{\bfr}) = \sum_{ m \sigma} C_{L  m, 1\sigma}^{JM}\,
  Y_{L  m}(\hat{\bfr})\, \bm{e}_{\sigma}\,,
\end{align}
and $\bm{e}_{\sigma}$ is the spherical component of the unity vector.

Inserting expansions (\ref{8}), (\ref{10}) and (\ref{11}) into Eq.~(\ref{1})
and employing the standard angular-momentum technique, we arrive at
\begin{widetext}
\begin{align} \label{14}
\bigl< \bfk \lambda| {\rho}_f |\bfk \lambda^{\prime}\bigr>
 =  &\ 8(2\pi)^4\, \sum_{\kappa_i \kappa_i^{\prime}
    \kappa_f} \sum_{LL^{\prime}\kappa g t}
  \sum_{\gamma_1 \gamma_2}
D^{g}_{\gamma_1\gamma_2}(\hat{\bfk})\,
  \rho^{(i)}_{\kappa,-\gamma_1}\, i^{l_i-l_i^{\prime}-L+L^{\prime}}\,
  e^{i\Delta_{\kappa_i}-i\Delta_{\kappa_i^{\prime}}}\,
 \left[L,L^{\prime},j_i,j_i^{\prime},l_i,l_i^{\prime},g,\kappa\right]^{1/2}\,
\nonumber \\ &\times
(-1)^{j_i^{\prime}-j_f+l_i+g+\kappa}\,
   C_{L^{\prime}\lambda^{\prime}, L-\lambda}^{g\gamma_2}\,
   C_{l_i0,l_i^{\prime}0}^{t0}\,C_{g-\gamma_1,\kappa\gamma_1}^{t0}\,
  \SixJ{L}{j_f}{j_i}{j_i^{\prime}}{g}{L^{\prime}}\,
  \NineJ{\nicefrac12}{\nicefrac12}{\kappa}{j_i^{\prime}}{j_i}{g}{l_i^{\prime}}{l_i}{t}
\nonumber \\ &\times
 \sum_{pp^{\prime}}
    (-i\lambda)^p\, (i\lambda^{\prime})^{p^{\prime}}\,
 \left< \vare_i\kappa_i\left|\left| \balpha\cdot \bm{a}^{(p)}_{L}\right|\right|\vare_f\kappa_f\right>^*\,
 \left< \vare_i\kappa_i^{\prime}\left|\left| \balpha\cdot
 \bm{a}^{(p^{\prime})}_{L^{\prime}}
               \right|\right|\vare_f\kappa_f\right>\,,
\end{align}
\end{widetext}
where $[x_1,x_2,\ldots] \equiv (2x_1+1)(2x_2+1)\ldots$, $l_a =
|\kappa_a+\nicefrac12|-\nicefrac12$,
 $l_a^{\prime} =
|\kappa_a-\nicefrac12|-\nicefrac12$, $j_a = |\kappa_a|-\nicefrac12$,
and the reduced matrix
elements are evaluated in Appendix~\ref{app:A}.

In this work, we will present our results in terms of the differential cross
section and the Stokes parameters of the emitted photon as functions of the polarization
vector of the incident electron: $\sigma(P_x,P_y,P_z)$, $P_1(P_x,P_y,P_z)$,
$P_2(P_x,P_y,P_z)$, and $P_3(P_x,P_y,P_z)$. We consider four
choices  of the polarization of the incident electron: (i) unpolarized,
$(P_x,P_y,P_z) = (0,0,0)$; (ii) polarized transversely within the
reaction plane, $(P_x,P_y,P_z) = (1,0,0)$; (iii) polarized perpendicularly to
the reaction plane $(P_x,P_y,P_z) = (0,1,0)$; and (iv)
longitudinally polarized, $(P_x,P_y,P_z) = (0,0,1)$. Not all of the polarization
correlations for these 4 choices are independent and nonvanishing. Properties of
the angular-momentum coefficients in Eq.~(\ref{14}) lead to the following
identities,
\begin{align} \label{15}
& P_1(0,0,0) = P_1(1,0,0) = P_1(0,0,1) \,,\\
& P_2(0,0,0) = P_2(0,1,0) = 0 \,, \\
& P_3(0,0,0) = P_3(0,1,0) = 0 \,, \\
& d\sigma(0,0,0) = d\sigma(1,0,0) = d\sigma(0,0,1)\,,
\end{align}
where $d\sigma(P_x,P_y,P_z) \equiv (k/Z^2)\, d\sigma(P_x,P_y,P_z)/(dk\,d\Omega_k)$.
Taking into account the above identities leaves
us with 8 independent quantities to be calculated:
$d\sigma(0,0,0)$, $d\sigma(0,1,0)$,
$P_1(0,0,0)$, $P_1(0,1,0)$, $P_2(1,0,0)$, $P_2(0,0,1)$, $P_3(1,0,0)$, and $P_3(0,0,1)$.

In previous calculations reported in the literature, the polarization
correlations were often parameterized in terms of the coefficients $C_{ij}$ introduced
by Tseng and Pratt \cite{tseng:73}.
In order to simplify comparison with previous studies, we give
the list of correspondence between the coefficients $C_{ij}$ and the present
notations,
\begin{align} \label{16}
&C_{03} =  P_1(0,0,0) \,,\\
&C_{11} = -P_2(1,0,0) \,,\\
&C_{12} = -P_3(1,0,0) \,,\\
&C_{23} = P_1(0,0,0)-P_1(0,1,0)\,,\\
&C_{31} = P_2(0,0,1)\,,\\
&C_{32} = P_3(0,0,1)\,,\\
&C_{20} =  1-\frac{d\sigma(0,1,0)}{d\sigma(0,0,0)}\,.
\label{16a}
\end{align}

\section{Numerical calculation}
\label{sec:numerics}

The problem of calculating the final-state density matrix (\ref{14}) is now reduced
to the evaluation of the radial integrals and the summation over the
angular-momentum and multipole quantum numbers. The radial integrals to be
evaluated are
\begin{eqnarray} \label{nu1}
J^{12}_{l}(a,b) & =&  \int_{0}^{\infty}
dr\,r^2\,g_{\vare_a,\kappa_a}(r)\,f_{\vare_b,\kappa_b}(r)\,  j_l(kr) \,,
  \\
J^{21}_{l}(a,b)  & =& \int_{0}^{\infty}
dr\,r^2\,f_{\vare_a,\kappa_a}(r)\,g_{\vare_b,\kappa_b}(r)\,  j_l(kr) \,,
 \label{nu1a}
\end{eqnarray}
where $g$ and $f$ are the radial components of the continuum Dirac wave
function and $j_l$ is the spherical Bessel function. A straightforward
numerical evaluation of these radial integrals is problematic since all
three functions in the integrand are highly oscillating and slowly decreasing
for large values of $r$.

Several methods were reported in the literature for
evaluation of such highly oscillatory integrals. For the case of
the point-Coulomb potential, the
integrals can be evaluated analytically in terms of the Appel functions of
three complex variables \cite{rozics:64,dugne:75}. Weak points of this
method are, first, the restriction to the particular choice of the atomic
potential and, second, the absence of reliable numerical methods for evaluation of
the Appel functions for a wide range of parameters and arguments.

The method
used in calculations by Tseng and Pratt \cite{tseng:71,tseng:73} is based on dividing
the integration region $(0,\infty)$ into two parts, the inner and the outer ones, in such a
way that the wave function in the outer region can be approximated by the
(phase-shifted) free-field solutions. In the inner part, the Dirac equation
is solved and the radial integrations are performed numerically. In the outer
region, the analytical form of the free-field solution is exploited and the
radial integrals are performed by the so-called integration by parts method,
which was never described in the literature \cite{tseng:70:unpublished}.

In the present work, the radial integrals are evaluated numerically after
rotating the integration contour in the complex $r$ plane.
For the first time this elegant method was used probably in Ref.~\cite{vincent:70}, in
connection with nuclear collision problems. For the case of the integrals involving
three spherical Bessel functions, this method was studied in detail in
Ref.~\cite{davies:88}. Integrals involving the nonrelativistic Coulomb
functions and negative powers of the radial coordinate were investigated 
in Ref.~\cite{sil:84}. More recenly, the complex-plane rotation method was
used in calculations of the $(e,2e)$ process \cite{keller:94} and the Compton
scattering \cite{bergstrom:93}. In the present work, we extend the complex-plane
integration method to the case of the Dirac Coulomb wave functions. The
calculational scheme is described in detail in Appendices \ref{app:B}
and \ref{app:C}. This method allowed us to evaluate the radial
integrals required up to typically 9-digit precision.

The continuum-state wave functions for the general case of the screening potential were
obtained by numerical solution of the Dirac equation with help of the RADIAL
package by Salvat et al.~\cite{salvat:95:cpc}. For the point-Coulomb
potential, we used the analytical representation for the Dirac eigenstates in
terms of the Whittaker functions, see Eqs.~(\ref{eq1}) and
(\ref{eq2}). The Whittaker $M$ and $W$ functions were
evaluated by a generalization of our codes developed in
Ref.~\cite{yerokhin:99:pra}.

In the present work, we perform calculations for two choices of the atomic
potential: (i) the point-nucleus Coulomb potential and (ii) the screening
potential of a neutral atom. The screening potential was constructed as
\begin{align} \label{nu2}
V_{\rm scr}(r) = V_{\rm nuc}(r) + \alpha \int_0^{R_0}dr'\,\frac{1}{\max(r,r')}\,\rho(r')\,,
\end{align}
where $V_{\rm nuc}$ is the potential of the (extended-size) nucleus, $R_0$
is the radius of the atom, and $\rho(r)$ is the radial electron density of atomic
orbitals normalized by
\begin{align} \label{nu3}
\int_0^{R_0}dr\,\rho(r) = Z\,.
\end{align}
The radial atomic electron density was calculated by the multiconfigurational
Dirac-Fock method by using the GRASP package \cite{parpia:96}. We note that the
previous calculations by Tseng and Pratt \cite{tseng:71,tseng:73} used a
somewhat simpler model
of the screening potential, which included only a local part of the exchange
interaction between the atomic electrons.

Once the radial integrals are successfully evaluated,
the main problem in numerical calculation is to ensure the convergence of the
multiple sum in Eq.~(\ref{14}). When all triangular selection rules are taken
into account, two summations remain unbound, which can be chosen to be $|\kappa_i|$
and $|\kappa_f|$. The convergence properties of the sum depend strongly both on the
kinetic energy of the incoming electron $E$ and on the fractional part of it
carried away by the photon, $k/E$. The convergence is fast for
$E<1$~MeV and close to the hard photon end of the spectrum (the tip region, $k/E\sim 1$)
and becomes rather slow for larger impact energies and(or) near the
soft photon end. In the tip region, the expansion over the final-state quantum
number $|\kappa_f|$ converges rapidly, so that a non-symmetric
partial-wave cutoff is appropriate. Our calculations in this region included
up to  $(n_i,n_f) = (80,10)$
partial waves. Further away from the tip region, we used the symmetric
configurations up to $(n_i,n_f) = (55,55)$ partial waves, with further
extension being complicated by numerical instabilities.

\section{Connection between bremsstrahlung and radiative recombination}
\label{sec:tip}

From the general physical point of view, it is natural to expect a relation
between the bremsstrahlung (BS) at the
hard photon end of the spectrum (the tip region), where the electron transfers
all its kinetic energy to the photon, and the radiative recombination
(RR) into highly excited atomic states.
This connection was studied previously by a number of authors
\cite{fano:59,lee:75,feng:83,jakubassa:10}. Approaching the tip region from
the continuum side was invesigated in Ref.~\cite{pratt:75}.
In this work, we would like to demonstrate
the connection between these two processes
by explicit numerical calculations.

For the Stokes parameters, the connection between RR and BS is most
transparent. It can be observed that, for RR into atomic states with a given angular
momentum quantum number $\kappa$, the Stokes parameters have a well-defined limit
as the principal quantum number $n\to\infty$. In order to make a connection to
BS, one has to account for the recombination into subshells with
different values of $\kappa$ (for a fixed value of
$n$) and then evaluate the limit of $n\to \infty$. More specifically, the RR Stokes
parameter $P_1$ is extrapolated to the continuum threshold as
\begin{align} \label{aa1}
 P_1^{\rm tip} &\ = \lim_{n\to \infty}
 \frac{\sum_{\kappa}\left[I_{0^{\circ}}(n\kappa)-I_{90^{\circ}}(n\kappa)\right]}
 {\sum_{\kappa}\left[I_{0^{\circ}}(n\kappa)+I_{90^{\circ}}(n\kappa)\right]}\,,
\end{align}
where $I_{\chi}(n \kappa)$ denotes the intensity of the photons emitted under
angle $\chi$ in
RR of an electron into a bound atomic substate $\ketm{n \kappa}$. For explicit
formulas and details on $I_{\chi}(n \kappa)$, we refer the reader to
the recent review by Eichler and St\"ohlker \cite{eichler:07:review}. The
other Stokes parameters $P_2$ and $P_3$ are obtained analogously to
Eq.~(\ref{aa1}), by using Eqs.~(\ref{4a}) and (\ref{7}).

In actual calculations, the limit $n\to\infty$ was evaluated by going to
sufficiently high values of $n$ (typically, $n\approx 15-20$), whereas the
summation over $\kappa$ was restricted to a few terms. (The
$s$, $p$, $d$, $f$, and $g$ waves were taken into account.)

A somewhat different procedure is
required for approaching the continuum threshold limit
in the case of the cross section. It is well-known that the cross section of RR into
the individual Rydberg states falls off as $\sim n^{-3}$. Making a connection to
the BS \cite{jakubassa:10}, one should take into account that the
individual levels cannot be resolved experimentally as $n\to\infty$. The
measured quantity depends on the resolution of the photon detector $\Delta k$
and is given by  
\begin{equation}
\left< \frac{d\sigma^{\rm RR}_{\kappa}}{d\Omega_k}  \right>_{\Delta k} =
\frac1{\Delta k}\, \sum_{n \geq n_0} \frac{d\sigma^{\rm RR}_{n\kappa}}{d\Omega_k}\,,
\end{equation}
where $\sigma^{\rm RR}_{n\kappa}$ is the cross section of RR into an individual level with
given $n$ and $\kappa$ and
$n_0$ is defined by the condition that the binding energy $E_{n_0} = \Delta
k$. In order to approach the continuum threshold, we 
sum over $\kappa$ and take the limit $n_0\to \infty$,
\begin{equation}
   \label{cross_section_SWL}
   \frac{{d}^2 \sigma^{\rm tip}}{{ d}k\, { d}\Omega_k} =
   \lim\limits_{n_0 \to \infty} \left( \frac{1}{\Delta k(n_0)} \,
   \sum\limits_{n \ge n_0} \, \sum\limits_{\kappa} \,
   \frac{{d} \sigma^{\rm RR}_{n\kappa}}{{ d}\Omega_k} \right) \, .
\end{equation}
In the limit $n\to\infty$, one can use the nonrelativistic expression for
the binding energy, $\Delta k(n) =  m(Z\alpha)^2/(2 n^2)$, and
the well--known asymptotic behavior of the RR cross section,
${d} \sigma^{\rm RR}_{n\kappa}/ {d}\Omega_k =
A_{\kappa}(\theta)/n^3$, with the result
\begin{equation}
   \label{cross_section_SWL_final}
   \frac{{ d}^2 \sigma^{\rm tip}}{{ d}k\, { d}\Omega_k} =
     \sum_{\kappa}\frac{A_{\kappa}(\theta)}{m(Z\alpha)^2} \, .
\end{equation}
In our calculations, we determine the parameter $A_{\kappa}(\theta)$ by fitting
the RR cross sections calculated for a series of $n$, to the $1/n^3$ scaling law.

\begin{figure*}
  \centerline{\includegraphics[width=\textwidth]{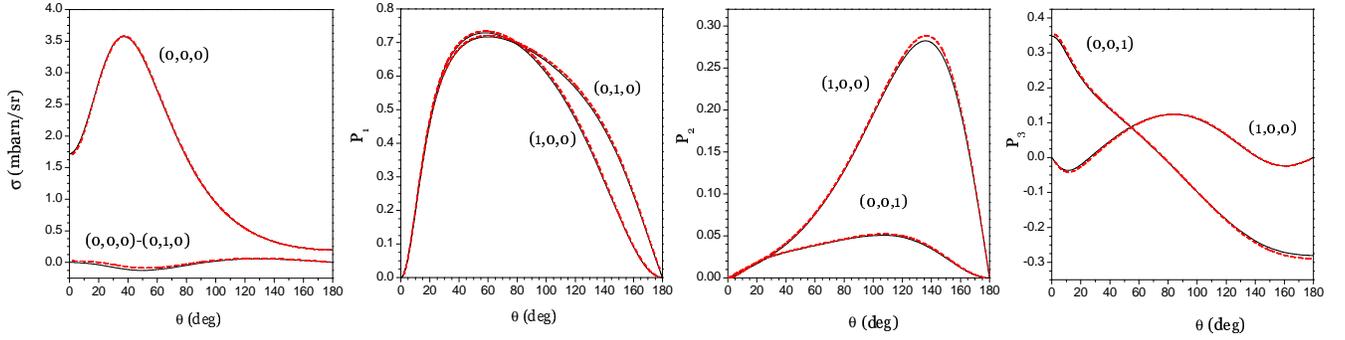}}
\caption{(Color online) Comparison of the bremsstrahlung
at the hard photon end of the spectrum (solid line, black) with the
continuum-threshold extrapolation of the radiative recombination (dashed line,
red), for the double differential cross section
$(k/Z^2)\, d\sigma/(dk\,d\Omega_k)$ and the Stokes parameters.
Calculations are performed for $Z=79$, $E = 100$~keV, and the point-nucleus Coulomb potential.
  \label{fig:tip} }
\end{figure*}

Fig.~\ref{fig:tip} illustrates the connection between BS and RR, as obtained in
our numerical calculations for the double differential cross section
and the Stokes parameters in the case of the point Coulomb potential.
We observe good agreement in all cases; the small remaining
deviation is attributed to extrapolation errors. It is important that our RR
and BS calculations are completely independent from each other. The BS
calculation is performed as described in this work, whereas the RR calculation
is carried out as reported in Ref.~\cite{surzhykov:03}, with the radial
integrations performed analytically. So, the agreement observed is also as
an important cross-check between two different calculational approaches.

\section{Results and discussion}
\label{sec:num}

\begin{figure*}
  \centerline{\includegraphics[width=\textwidth]{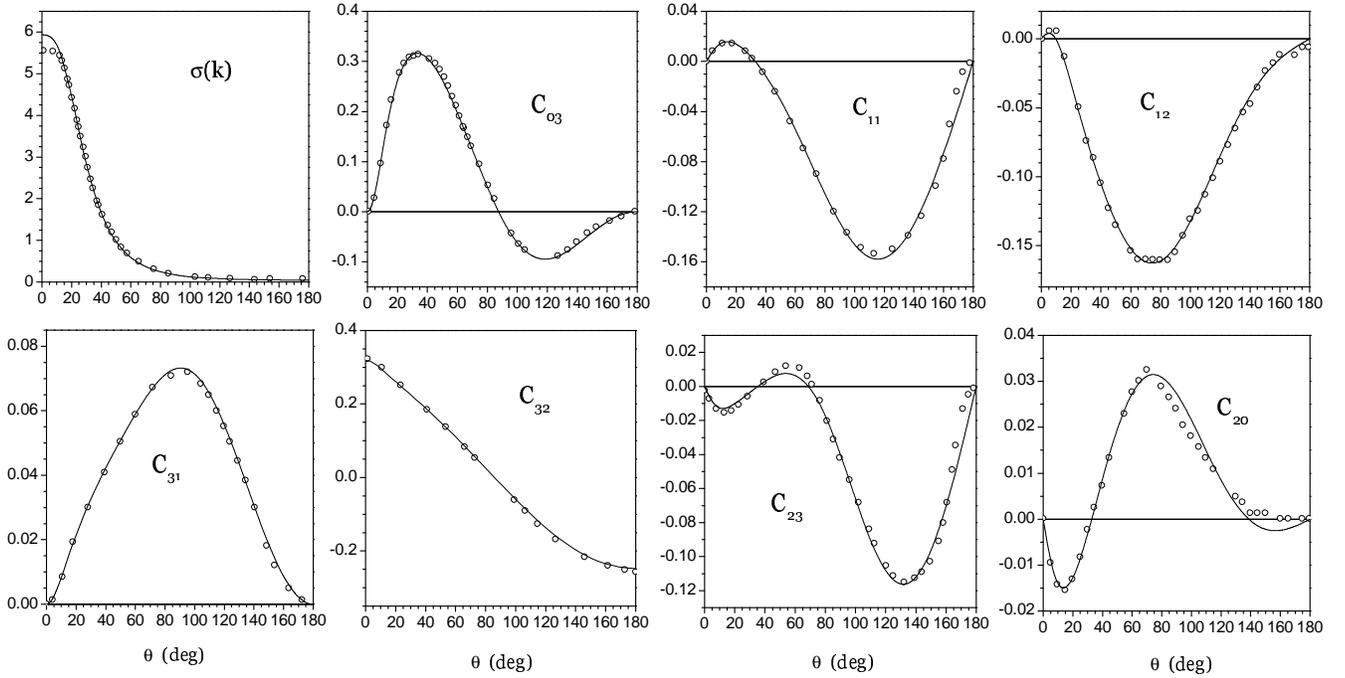}}
\caption{Comparison of the present results (solid line) with those by
Tseng and Pratt \cite{tseng:71,tseng:73} (open circles) for the double differential cross
section $(k/Z^2)\, d\sigma/(dk\,d\Omega_k)$ for the initially unpolarized electrons, in
mbarn/sr, and for the polarization correlations $C_{ij}$.
The calculations are performed for $Z=79$,
$E = 500$~keV, $k = 250$~keV, and the point-nucleus Coulomb potential.
  \label{fig:comparison} }
\end{figure*}

We begin this section by comparing our numerical results with those obtained in the
best previous partial-wave expansion calculations by Tseng and Pratt
\cite{tseng:71,tseng:73}. Table~\ref{tab:compar} presents comparison for the
bremsstrahlung cross section $\sigma(k)$ given by Eq.~(\ref{3}) for the gold
target, both for the Coulomb and screening potentials. We observe
excellent agreement between the calculations performed for the Coulomb potential. For
the screening potential, all our results are 
by about 1\% smaller than the ones by Tseng and Pratt. This
difference is apparently
due to a more realistic screening potential used in this work [see the discussion
after Eq.~(\ref{nu3})]. Results of the
two calculations for the double differential cross section and for all
independent polarization corrections $C_{ij}$ [see Eqs.~(\ref{16})-(\ref{16a})] are
compared in Fig.~\ref{fig:comparison} for the case of the Coulomb potential.
We find nearly perfect agreement in all cases. As a matter of fact, the
agreement observed is quite remarkable, taking into account that the
calculation by Tseng and Pratt was accomplished four decades ago.

\begin{figure*}
  \centerline{\includegraphics[width=0.8\textwidth]{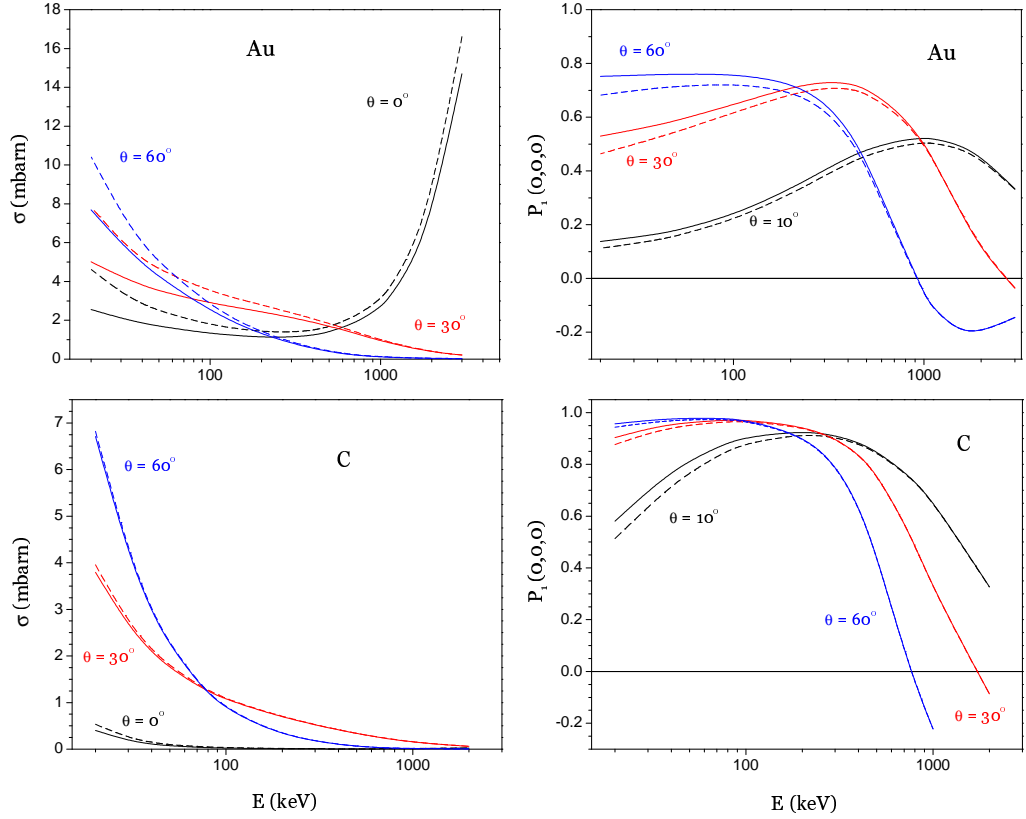}}
\caption{(Color online) Double differential cross section $(k/Z^2)\,
d\sigma/(dk\,d\Omega_k)$ and the Stokes parameter $P_1$ for several
angles of the emitted photon $\theta$, as functions of the kinetic
energy of the incoming electron $E$, for gold (upper row) and carbon (lower
row) targets, for the Coulomb (dashed line) and screened (solid line)
potentials. For all graphs, the kinetic energy of the final electron is $E_f =
1$~keV and initially unpolarized electrons are taken.
  \label{fig:screening} }
\end{figure*}

In Fig.~\ref{fig:screening}, we study the effect of the screening on the cross
section and the Stokes parameter $P_1$ for the initially unpolarized
electrons. First of all, we observe that the sign
of the effect varies: the screening
reduces the cross section but increases the Stokes parameters. For the cross
section, the screening effect depends strongly on the nuclear charge $Z$. The effect is
barely recognizable (in the region of energies studied) for carbon, while for
gold, it is significant for all energies.

It is interesting to observe that the bremsstrahlung on the bare nucleus and
the neutral atom are very much alike for not-too-small impact energies. This
is explained by the fact that the dominant contribution to the matrix element
of the amplitude of the process comes from the small distances of
configuration space. At these distances, the continuum-state electron
``feels'' mainly the bare nuclear charge and the screening effect is
relatively small. At larger distances, the continuum-state electron wave
function oscillates rapidly, so that the contribution to the radial integal is
small. When the impact energy increases, the region of
configuration space responsible for the dominant part of the integral
shifts towards the nucleus and the screening effect decreases. 

For $E=20$~keV and gold target, the screening reduces the
forward-scattering cross section
by more than a factor of 2, whereas at energies of a few MeV, it is still
a 10\% effect. For the Stokes parameters, the screening effect is smaller than
for the cross section and stays within
10\% even at the lowest energy studied, $E=20$~keV. We observe that for
energies below 20~keV and gold target, calculations of the cross section within the
screening-potential approximation rapidly become meaningless as the screening
effect tends to grow very fast as the energy decreases. On the contrary, the Stokes
parameters turn out to be much less sensitive to the electronic structure of the
target. We also find that for the high energies above 1~MeV, the screening
effect is significant mainly for the cross section at close-to-zero angles,
where it amounts to about 10\%.

\begin{figure*}
  \centerline{\includegraphics[width=0.8\textwidth]{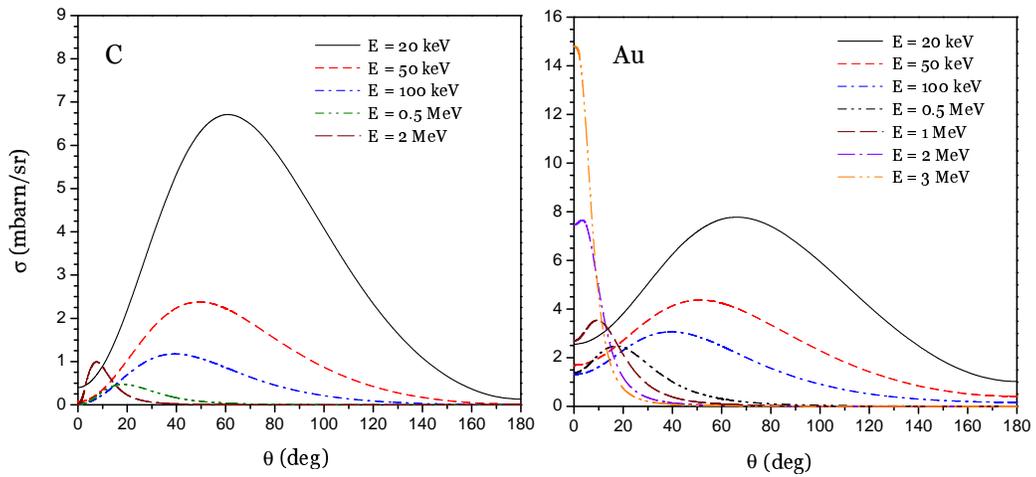}}
\caption{(Color online) Double differential cross section $(k/Z^2)\,
d\sigma/(dk\,d\Omega_k)$ for different values of the kinetic
energy of the incoming electron $E$, for neutral carbon (left) and gold (right)
targets. Kinetic energy of the final electron is $E_f =
1$~keV and initially unpolarized electrons are taken.
  \label{fig:dsc} }
\end{figure*}

It is interesting to observe that the dependence of the forward-scattering cross section
on the impact energy in the tip region 
is drastically different for the low-$Z$ (carbon) and high-$Z$ (gold) targets. This
difference is studied in more detail in Fig.~\ref{fig:dsc}. We observe that
for carbon, the forward-scattering cross section at the tip nearly vanishes for large
range of impact energies, as could be expected from the nonrelativistic
theory. On the contrary, the gold
target corresponds to the highly relativistic regime and its
forward-scattering cross section is strongly non-zero for all energies. At high impact
energies $E>0.5$~MeV, the forward-scattering cross section grows rapidly
with the increase of $E$, whereas the angular distrubution of the
cross section becomes localized at the increasingly smaller regions near 
the forward direction.

\begin{figure*}
  \centerline{\includegraphics[width=0.8\textwidth]{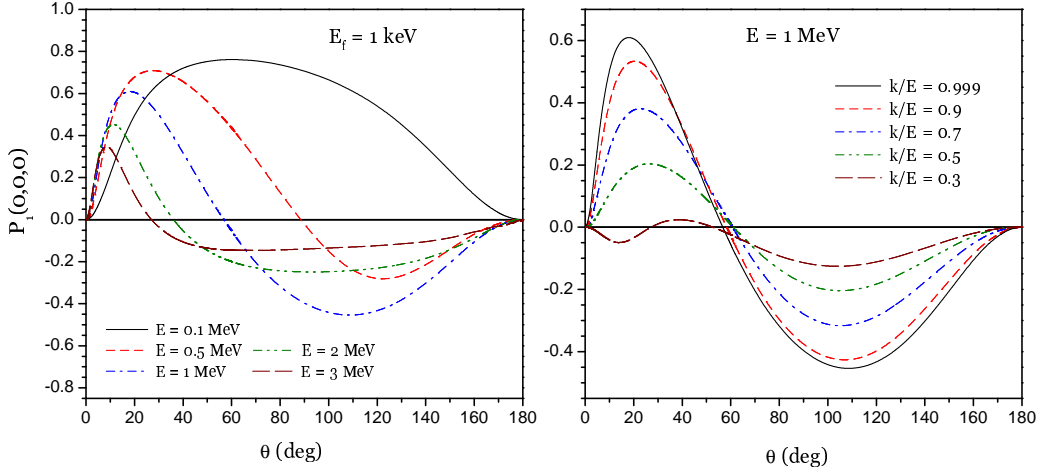}}
\caption{(Color online) Stokes parameter $P_1$ for the neutral gold target
and initially unpolarized electrons. Left panel contains plots for 
different values of the kinetic energy of the incoming electron $E$ and the
kinetic energy of the final-state electron fixed by $E_f = 1$~keV. In right
panel, the initial energy is fixed by $E = 1$~MeV and the final-state energy
is varied.
  \label{fig:P1000}
}
\end{figure*}

In Fig.~\ref{fig:P1000}, we study the energy dependence of the Stokes
parameter $P_1$, the only nonvanishing Stokes parameter for the initially
unpolarized electrons. We observe that the shape of the angular distribution of
$P_1$ depends strongly on the energy of the initial electron $E$. For small
$E$, $P_1$ tends to its nonrelativistic value of $P_1 = 1$, whereas for the
increasingly larger energies, maximum of the angular distribution decreases
and shifts towards the forward direction and its tail crosses the zero axis (the
``crossover'' feature \cite{tseng:73}). On the contrary, when the initial
energy $E$ is fixed and the energy of the final electron is varied, the shape
of the angular distribution stays much the same but its amplitude decreases.

\begin{figure*}
  \centerline{\includegraphics[width=0.8\textwidth]{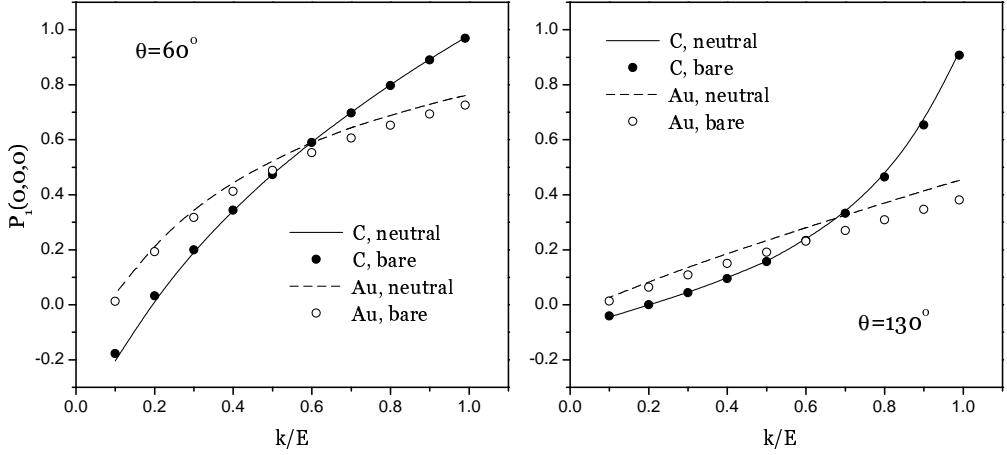}}
\caption{Stokes parameter $P_1$ for the initially unpolarized electrons with
$E = 100$~keV, for
different values of the energy $k$ and the angle $\theta$
of the emitted photon, for gold and carbon targets and
for the Coulomb and screening potentials. 
  \label{fig:P1AuC}
}
\end{figure*}

In Fig.~\ref{fig:P1AuC}, we plot the Stokes parameter $P_1$
as a function of the fractional energy carried away by
the photon, $k/E$, for two targets, carbon and gold, and for two values of the
angle of the emitted photon. These plots correspond to the experiment that is
currently underway in TU Darmstadt \cite{maertin:CAARI}.

\begin{figure*}
  \centerline{\includegraphics[width=0.8\textwidth]{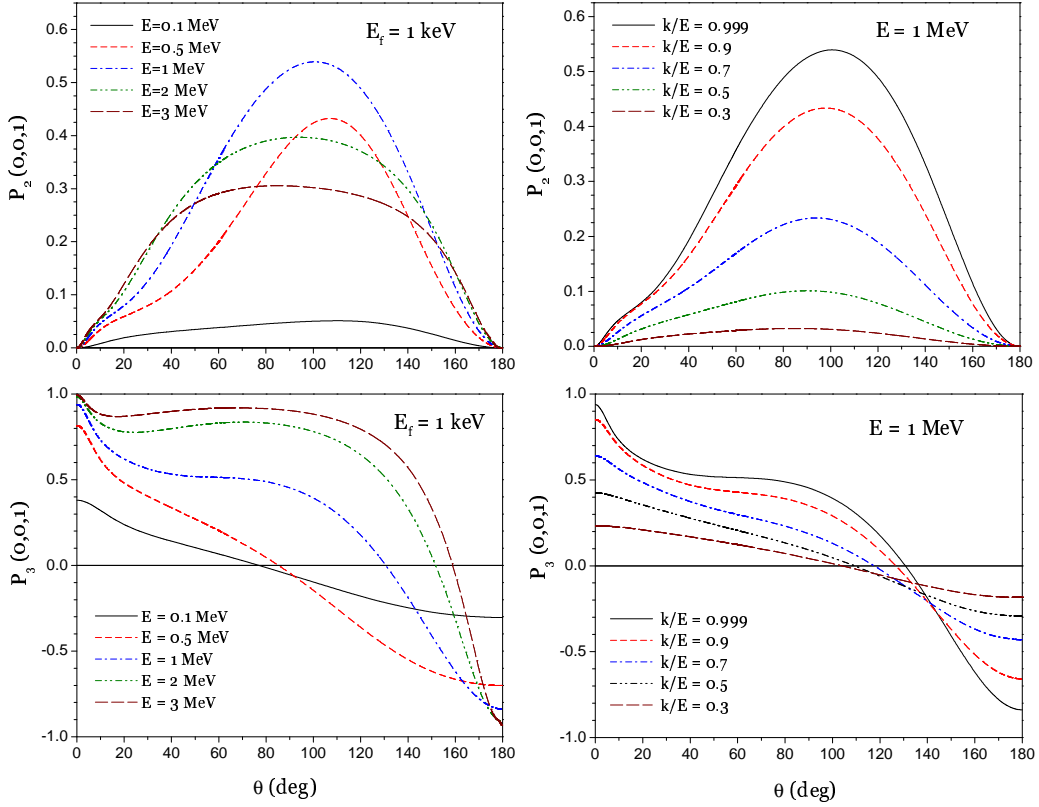}}
\caption{(Color online) The same as Fig.~\ref{fig:P1000} but for the Stokes parameters $P_2$
  (upper row) and $P_3$ (lower row) and for the initially longitudinally
  polarized electrons.
  \label{fig:P23001} }
\end{figure*}

We now turn to the case of the initially polarized electrons. In the present
investigation, we restrict
ourself to studying the longitudinal polarization. Fig.~\ref{fig:P23001}
presents our results for the Stokes parameters $P_2$ and $P_3$ as functions of
the initial energy $E$ and of the energy of the final-state electron. (We
recall that $P_1$ for the longitudinally polarized electrons is the same as for
the unpolarized ones and shown in Fig.~\ref{fig:P1000}.) The
Stokes parameter $P_2$ is of pure relativistic origin and thus is very small for small
initial energies. With increase of $E$, $P_2$ becomes comparable to $P_1$, has its maximum
at about $E=1$~MeV, and then gradually descreases. The third Stokes parameter
$P_3$ is also a relativistic effect and thus vanishes for small initial
energies. However, in the high-energy region and $k/E\sim 1$, $P_3$ changes
its behaviour drastically and approaches unity everywhere except for the 
backward direction.

\section{Summary}

In this paper, we report a detailed study of the electron-atom
bremsstrahlung process within the rigorous relativistic approach based on the
partial-wave expansion of the Dirac wave functions in the external atomic field.
Assuming that the final-state electron is not observed, we evaluate the
double-differential cross section and all polarization correlations. Unlike in
the previous studies, our description of the polarization correlations is
formulated entirely in terms of the Stokes parameters, which are directly
related to quantities observed in modern experiments. For our calculations,
we developed an efficient and reliable scheme of evaluation of the radial
integrals for the free-free transitions, based on the complex-plane rotation of
the integration contour. The method is applied for the Dirac solutions in both the
point-Coulomb potential and the finite-range screening potential.

The numerical procedure was carefully checked by comparing our results against those
reported in the literature. Comparison of our results obtained
for the hard-photon end point of the bremsstrahlung spectrum 
with the extrapolation of the radiative recombination
results yielded a numerical proof of the connection between bremsstrahlung and
radiative recombination and served as an additional check of our computational
scheme.

The numerical results reported present a detailed analysis of (i)
the screening effect induced by the electrons of the target on the cross
section and polarization correlations and
(ii) the energy dependence of the polarization
correlations,
with the main focus on the high-energy region, which is of primary interest
in the future experiments at GSI. We conclude that the tip region of the
bremsstrahlung spectrum is the most appropriate for studying the
polarization correlations, as all polarization correlations have their maximum
values there.

\section*{Acknowledgement}

The authors are grateful to Prof.~R.~Pratt for valuable and interesting
discussions. Help of Dr.~L.~Sharma in using the GRASP package is acknowledged.
The work reported in this paper was supported by the Helmholtz Gemeinschaft
(Nachwuchsgruppe VH-NG-421).


\begin{table}
\caption{
Comparison of the results of the present calculation
of the bremsstrahlung cross section $\sigma(k)$
with that by Tseng and Pratt
\cite{tseng:71}, for the gold target ($Z=79$) and
the Coulomb and screening potentials, in mbarn.
$E$ is the kinetic energy of the incoming electron and $k$ is the energy of
the emitted photon.
\label{tab:compar}}
\begin{ruledtabular}
  \begin{tabular}{clcccc}
$E$  & \multicolumn{1}{c}{$k/E$}  &  This work & Ref.~\cite{tseng:71}&  This
    work & Ref.~\cite{tseng:71} \\
 \multicolumn{1}{c}{[MeV]}     &              & \multicolumn{2}{c}{Coulomb} & \multicolumn{2}{c}{screened}\\
    \hline\\[-5pt]
   0.050   &  0.6         &     42.61   &  42.59 &   35.06   &  35.62 \\
           &  0.4         &     46.74   &  46.72 &   36.92   &  37.29 \\
   0.180   &  0.6         &     14.68   &  14.67 &   13.22   &  13.34 \\
   0.380   &  0.6         &     8.555   &  8.522 &   7.884   &  7.942 \\
   0.500   &  0.96        &     4.791   &  4.789 &   4.463   &  4.526 \\
           &  0.5         &     8.201   &  8.142 &   7.562   &  7.600
  \end{tabular}
\end{ruledtabular}
\end{table}

%
%

%
%

%
%

%
%

%
%


%
%

%
%


%
%
%
\appendix

\begin{widetext}
\section{Reduced matrix elements}
\label{app:A}

Results for the reduced matrix elements in Eq.~(\ref{14}) are
\begin{align}
(-i) \bigl< \vare_a\kappa_a|| \balpha\cdot  \bm{a}^{(0)}_{l}||\vare_b\kappa_b\bigr>
 = &\  J^{12}_{l}(a,b)\, s_{ll}(\kappa_a,-\kappa_b)
- J^{21}_{l}(a,b)\, s_{ll}(-\kappa_a,\kappa_b)
\,,
\end{align}
\begin{align}
(-i) \bigl< \vare_a\kappa_a|| \balpha\cdot  \bm{a}^{(1)}_{l}||\vare_b\kappa_b\bigr>
 = &\
 \sqrt{\frac{l+1}{2l+1}} \biggl[
 J^{12}_{l-1}(a,b)\, s_{l\,l-1}( \kappa_a,-\kappa_b)
-  J^{21}_{l-1}(a,b)\,
s_{l\,l-1}(-\kappa_a,\kappa_b) \biggr]
 \nonumber \\
&- \sqrt{\frac{l}{2l+1}} \biggl[
 J^{12}_{l+1}(a,b)\, s_{l\,l+1}( \kappa_a,-\kappa_b)
-  J^{21}_{l+1}(a,b)\,
s_{l\,l+1}(-\kappa_a,\kappa_b) \biggr]
\,,
\end{align}
where the angular coefficients are given by
\begin{align}
s_{LJ}(\kappa_1,\kappa_2) = \bigl<\kappa_1|| \bsigma  \cdot \bm{Y}_{LJ} ||
\kappa_2\bigr>
= \bigl< \kappa_1|| [Y_J \otimes \bsigma]_L  ||\kappa_2 \bigr>
 = \sqrt{\frac{3}{2\pi}}\,(-1)^{l_2}\,\bigl[j_1,j_2,l_1,l_2,L\bigr]^{1/2}\,
  C_{l_1 0, l_2 0}^{J 0}\,
  \NineJ{j_1}{l_1}{\nicefrac12}{j_2}{l_2}{\nicefrac12}{L}{J}{1}\,,
\end{align}
\end{widetext}
$\bm{Y}_{JLM}$ are the vector spherical harmonics defined by Eq.~(\ref{13})
and the radial integrals $J^{12}_l(a,b)$ and $J^{21}_l(a,b)$ are defined by
Eqs.~(\ref{nu1}) and (\ref{nu1a}).

\section{Free-free integrals: Coulomb case}
\label{app:B}

We need the radial integrals of the form
\begin{align} \label{eqrad}
J^{ij}_l(\vare_1,\kappa_1,\vare_2,\kappa_2) =
 \int_{0}^{\infty}
 dr\,r^2\,f_{\vare_1,\kappa_1,i}(r)\,f_{\vare_2,\kappa_2,j}(r)\,
   j_l(kr)\,,
\end{align}
where $j_l$ is the spherical Bessel function, $k = |\vare_1-\vare_2|$,
and $f_{\vare,\kappa,1} \equiv g_{\vare,\kappa}$
and  $f_{\vare,\kappa,2} \equiv f_{\vare,\kappa}$ are the upper and the
lower radial components of the Dirac wave function. In the following, we will
assume that $\vare_1>\vare_2>m$, which entails that $p_1-p_2-k>0$, where $p_i
= \sqrt{\vare_i^2-m^2}$. By this restriction, we exclude the possibility
$\vare_1 = \vare_2$, in which case
the method described in this section is not applicable.

For $\vare>m$, the radial components of the Dirac-Coulomb wave functions
normalized on the energy scale are given by \cite{eichler:95:book}:
\begin{align} \label{eq1}
g_{\vare,\kappa}(r) = &\
  N_{\kappa}\,\sqrt{\vare+1}\, (2pr)^{-3/2}\, {\rm Re}\,
 \biggl\{
 e^{i\bigl[\delta_{\kappa}-\frac{\pi}{2}\left(\gamma+\frac12\right)\bigr]}\,
 \nonumber \\ & \times
(\gamma+  i\eta)\,  M_{-1/2-i\eta,\gamma}(2ipr)\biggr\}\,, \\
f_{\vare,\kappa}(r) = &\
  -N_{\kappa}\,\sqrt{\vare-1}\, (2pr)^{-3/2}\, {\rm Im}\,
 \biggl\{
 e^{i\bigl[\delta_{\kappa}-\frac{\pi}{2}\left(\gamma+\frac12\right)\bigr]}\,
 \nonumber \\ & \times
(\gamma+
 i\eta)\,  M_{-1/2-i\eta,\gamma}(2ipr)\biggr\}\,,
\label{eq2}
\end{align}
where $p = \sqrt{\vare^2-m^2}$, $\eta = \Za\,\vare/p$, $\gamma =
\sqrt{\kappa^2-(\Za)^2}$,
\begin{align}
N_{\kappa} = 2\sqrt{\frac{p}{\pi}}\,e^{\pi\eta/2}\,\frac{\left|
  \Gamma(\gamma+i\eta)\right|} {\Gamma(2\gamma+1)}\,,
\end{align}
\begin{align}
e^{2i\delta_{\kappa}} = \frac{-\kappa+i\eta/\vare}{\gamma+i\eta}\,,
\end{align}
and $M_{\alpha,\beta}$ is the Whittaker function of the first kind
\cite{gradshteyn}.

It is clear that the integrals $J^{ij}_l$ can be expressed in terms of the
integrals involving two Whittaker functions and a Bessel function,
\begin{align} \label{3ms}
I_{\alpha_1,\gamma_1,\alpha_2,\gamma_2,l}(p_1,p_2,k) = &\
 \int_0^{\infty}dr\,r^{-1}\,
M_{\alpha_1,\gamma_1}(2ip_1r)\,
 \nonumber \\ & \times
 M_{\alpha_2,\gamma_2}(2ip_2r)\,j_l(kr)\,,
\end{align}
with the momenta satisfying the condition $p_1-p_2-k>0$. The integrand is
highly oscillatory
and slowly decreasing function for large values of $r$, so a straightforward
numerical evaluation of this integral up to a high accuracy is practically impossible.
The method used in the present work \cite{davies:88} is based on
the analytical continuation of the integrand into the complex $r$ plane. More
specifically, the integrand is separated into two parts which decrease
exponentially in the upper or lower half of the complex plane. By appropriate
rotations of the integration contour, the original oscillating integrand can be
converted into two smoothly decreasing ones. The resulting integrals can be
easily calculated by Gauss-Legendre quadratures up to typically
9 digit accuracy.

To realize this algorithm, we represent the first Whittaker function
in the right-hand-side of
Eq.~(\ref{3ms}) in terms of the Whittaker functions of the second kind
\cite{gradshteyn}
\begin{align} \label{m2w}
M_{\alpha,\beta}(z) &\ = \frac{\Gamma(2\beta+1)}{\Gamma(\beta-\alpha+1/2)}\,
 e^{i\pi s \alpha} \, W_{-\alpha,\beta}(-z)
\nonumber \\ &
+
\frac{\Gamma(2\beta+1)}{\Gamma(\beta+\alpha+1/2)}\,
 e^{i\pi s (\alpha-\beta-1/2)} \, W_{\alpha,\beta}(z)\,,
\end{align}
where $s = 1$ if ${\rm Im}(z)<0$ and $-1$ otherwise.
For $r = R+iz$, the asymptotic behavior of the Whittaker and Bessel functions
(for $p$ and $k>0$ and $|z|\to \infty$) is
\begin{align}
M_{\alpha,\beta}(2ipr) \sim e^{p|z|}\,, \ \
W_{\alpha,\beta}(2ipr) \sim e^{pz}\,, \ \
j_l(kr) \sim e^{k|z|}\,,
\end{align}
where only the leading exponential behavior is kept. These results,
together with the condition on the momenta, $p_1>p_2+k$,
show that the representation (\ref{m2w}) applied to the Whittaker function
with the largest momenta
$M_{\alpha_1,\gamma_1}(2ip_1r)$, splits the integrand of Eq.~(\ref{3ms}) into
two parts, one of which [with $W_{-\alpha_1,\gamma_1}(-2ip_1r)$] is regular in
the upper half of the complex $r$ plane and the other [with
$W_{\alpha_1,\gamma_1}(2ip_1r)$] is regular into the lower half-plane.
In both cases, the resulting integrand falls off as $\sim
e^{-(p_1-p_2-k)|z|}$, which makes possible accurate numerical evaluation of
the corresponding integrals.

In actual calculations, one should keep in mind that Eq.~(\ref{m2w}) represents
the regular (for small $z$) function $M$ in terms of the irregular functions
$W$. Because of this, it is advantageous to integrate Eq.~(\ref{3ms}) along the real
axis up to a certain value of $r = R$ and to perform
the rotation of the integration contour for $r>R$. So,
Eq.~(\ref{3ms}) is represented as
\begin{align}
I = \int_0^R dr\,I(r)+ \int_{0}^{\infty} dz\,i\,\bigl[I_+(R+iz)-I_-(R-iz)\bigr]\,,
\end{align}
where $I(r)$ stands for the integrand of the right-hand-side of
Eq.~(\ref{3ms}), and $I_+(r)$ and $I_-(r)$ are the parts of $I(r)$ regular in
the upper and lower half-plane, respectively, $I(r) = I_+(r)+I_-(r)$.

\section{Free-free integrals: neutral atom}
\label{app:C}

We now consider
the potential in the Dirac equation to be a finite-range screening potential
(having in mind a neutral atomic system),
\begin{align}
V(r) = V_{\rm scr}(r)\,,
\end{align}
where $ V_{\rm scr}(r) = 0$ for $r\geq R_0$. In this case,
in the outer region $r\geq R_0$, the solutions of the Dirac equation can be expressed in terms
of the free Dirac eigenfunctions, given by
\begin{align}
g_{\vare,\kappa}^{(0)}(r) &\ =
 \sqrt{\frac{p\,(\vare+1)}{\pi}}\,
 \biggl[ j_l(pr)\, \cos\delta - y_l(pr)\,\sin\delta \biggr]\,, \\
f_{\vare,\kappa}^{(0)}(r) &\ =
\frac{|\kappa|}{\kappa}\,\sqrt{\frac{p\,(\vare-1)}{\pi}}\,
 \biggl[ j_{\overline{l}}(pr)\, \cos\delta - y_{\overline{l}}(pr)\,\sin\delta \biggr]\,,
\end{align}
where $j_l$ and $y_l$ are the spherical Bessel and Neumann functions,
respectively; $l = |\kappa+1/2|-1/2$, $\overline{l} = |\kappa-1/2|-1/2$,
and $\delta$ is the scattering phase shift induced by the finite-range potential $V_{\rm
  scr}$ and determined by the matching procedure at $r = R_0$.
(In the absence of any field, $\delta = 0$.)
Note that the definition of the overall sign in the wave functions agrees with
that of Ref.~\cite{eichler:95:book}.
The definition used in Ref.~\cite{salvat:95:cpc} differs from the present one by
a factor of $-\kappa/|\kappa|$.

For the evaluation of radial integrals, it is convenient to express the free
Dirac eigenfunctions
in terms of the spherical Hankel functions of the first and second kind
($h^{(1)}_l$ and $h^{(2)}_l$, respectively),
\begin{align} \label{eq01}
g_{\vare,\kappa}^{(0)}(r) &\ =
 \sqrt{\frac{p\,(\vare+1)}{\pi}}\,\frac12\,
 \biggl[ h^{(1)}_l(pr)\, e^{i\delta} + h^{(2)}_l(pr)\,e^{-i\delta} \biggr]\,, \\
f_{\vare,\kappa}^{(0)}(r) &\ =
\frac{|\kappa|}{\kappa}\,\sqrt{\frac{p\,(\vare-1)}{\pi}}\,\frac12\,
 \biggl[ h^{(1)}_{\overline{l}}(pr)\, e^{i\delta} + h^{(2)}_{\overline{l}}(pr)\,e^{-i\delta} \biggr]\,.
  \label{eq02}
\end{align}
Taking into account that $h^{(1)}_l(pr) \sim e^{ipr}$ and $h^{(2)}_l(pr) \sim
e^{-ipr}$  as $r \to \infty$, we conclude that the first terms in the brackets of
Eqs.~(\ref{eq01}) and (\ref{eq02}) are regular in the upper half of the
complex $r$ plane, whereas the second terms are regular in the lower
half-plane.

The integral involving the free Dirac eigenfunctions can be
evaluated as
\begin{widetext}
\begin{align}
J^{ij}_l(\vare_1,\kappa_1,\vare_2,\kappa_2,R) &\ =
 \int_{R}^{\infty}
 dr\,r^2\,f^{(0)}_{\vare_1,\kappa_1,i}(r)\,f^{(0)}_{\vare_2,\kappa_2,j}(r)\,   j_l(kr)
\nonumber \\
&\ =
 \int_{0}^{\infty}
 dz\,
i\, \left[
x^2\,f^{(0),+}_{\vare_1,\kappa_1,i}(x)\,f^{(0)}_{\vare_2,\kappa_2.j}(x)\,j_l(kx)
-
{x^*}^2\,f^{(0),-}_{\vare_1,\kappa_1,i}(x^*)\,f^{(0)}_{\vare_2,\kappa_2,j}(x^*)\,j_l(kx^*)
\right]\,,
\end{align}
\end{widetext}
where $x = R+iz$, and superscripts ``$+$'' and ``$-$'' label the parts
of the eigenfunction
that are regular in the upper and lower half-plane, respectively. Using the
property of the spherical Hankel functions
\begin{align}
\left[ h^{(1)}_l(z)\right]^* = h^{(2)}_l(z^*)\,,
\end{align}
the expression for $J^{ij}_R$ is reduced to
\begin{align}
J^{ij}_l(\vare_1,\kappa_1,\vare_2,\kappa_2,R) &\ =
-2\, {\rm Im}\, \int_{0}^{\infty}
 dz\, x^2\,
\nonumber \\ & \times
f^{(0),+}_{\vare_1,\kappa_1,i}(x)\,f^{(0)}_{\vare_2,\kappa_2,j}(x)\,j_l(kx)\,.
\end{align}
It can be easily seen that the integrand in the above expression
falls off as $e^{-(p_1-p_2-k)z}$ for large $z$.

For $r$ smaller than the radius of the atom $R_0$, the wave functions are
obtained numerically by solving the Dirac equation, so this part of the
radial integration has to be performed along the real axis. When the energy of the
incoming electron becomes large, the integration within the radius of the atom
(which is of order of several atomic units) might become troublesome as the
integrand is rapidly oscillating. To simplify the numerical evaluation,
we exploit the fact that, for
heavy atoms, the maximum of the screening potential $V_{\rm scr}$ is localized close to the
nucleus. Further away from the maximum,
the numerical solutions of the Dirac equation resemble
the free asymptotic solutions, so that the difference between
them is a smooth, rapidly decreasing function.
We thus represent the radial integrals $J^{ij}_l$
[Eq.~(\ref{eqrad})] as
\begin{widetext}
\begin{align}
J^{ij}_l(\vare_1,\kappa_1,\vare_2,\kappa_2) &\ =
 \int_{0}^{R} dr\,r^2\,f_{\vare_1,\kappa_1,i}(r)\,f_{\vare_2,\kappa_2,j}(r)\, j_l(kr)
  \nonumber \\ &
+ \int_{R}^{R_0}
 dr\,r^2\,\biggl[f_{\vare_1,\kappa_1,i}(r)\,f_{\vare_2,\kappa_2,j}(r)-
   f^{(0)}_{\vare_1,\kappa_1,i}(r)\,f^{(0)}_{\vare_2,\kappa_2,j}(r)\biggr]\, j_l(kr)
  \nonumber \\ &
-2\, {\rm Im}\, \int_{0}^{\infty}
 dz\, x^2\,f^{(0),+}_{\vare_1,\kappa_1,i}(x)\,f^{(0)}_{\vare_2,\kappa_2,j}(x)\,j_l(kx)\,,
\end{align}
where $x = R+iz$. The free parameter $R<R_0$ is chosen as the smallest
distance for which the exact Dirac solution $f_{\vare,\kappa,i}(r)$ resembles
the free asymptotic solution $f^{(0)}_{\vare,\kappa,i}(r)$.
\end{widetext}

\end{document}